\documentstyle[12pt]{article}
\newcommand {\nc} {\newcommand}

  \nc \hel {${}^3 {\rm He}$}
 \nc \ba {\begin{eqnarray}}    \nc \ea {\end{eqnarray}}
 \nc \bc {\begin{center}}      \nc \ec {\end{center}}
	\nc	{\eq}[1] {Eq.~(\ref{#1})}
  \nc {\ra} {R}                   \nc {\ia} {I}

  \nc {\rc}[1] {\ra_#1}        \nc {\ic}[1] {\ia_#1}

  \nc {\cfa}[1] {a_{#1}}       \nc {\cfb}[1] {b_{#1}}
  \nc {\A}[1] {\mbox{$A_{#1}$}} 
  \nc {\ppa}[1] {\mbox{$PP'A_{_{#1}}$}} 

  \nc {\gp} {\\[1ex]}             \nc {\gap} {\\[2ex]}
  \nc {\f}  {\frac}               \nc {\s} {\sqrt}
  \nc {\q}  {\quad}               \nc {\qq} {\qquad}
  \nc {\lb} {\left[\,}            \nc {\rb} {\,\right]}
  \nc {\lp} {\left(}              \nc {\rp} {\right)}
  \nc {\m} {\mbox}
  \nc {\delsigL}        {\Delta\sigma_L}
  \nc {\delsigT}        {\Delta\sigma_T}
  \nc {\delL}           {\Delta_L}
  \nc {\delT}           {\Delta_T}
  \nc {\dsigdt}         {\frac{d\sigma}{dt}}
  \nc {\half}           {\mbox{$\frac{1}{2}$}}
  \nc {\ph}             {\phantom}
  \nc {\sigtot}         {\sigma_{\text{tot\/}}}
  \nc {\slope}          {B}

\nc{\re} {\mathop{\mathrm{Re}}} \nc{\im} {\mathop{\mathrm{Im}}} 
\nc{\case}[2] {\mbox{$\frac{#1}{#2}$}} 

  \nc {\kap} {\kappa}           
  \nc {\lama} {\half \kappa}
  \nc {\amp}[1] {\phi_{#1}}
  \nc {\aml}[2] {\phi_{#1}^{\mathrm{#2}}}
  \nc {\text}[1] {\mbox{\scriptsize{#1}}}            

\def\Journal#1#2#3#4{{#1} {\bf #2}, #3 (#4)}

\def\PRD{{Phys. Rev.} D}
\def\PR{{Phys. Rev.}}

\def\SJNP{Sov.\ J.\ Nucl.\ Phys.}

\newcommand{\be}{\begin{equation}}
\newcommand{\ee}{\end{equation}}
\newcommand{\bea}{\begin{eqnarray}}
\newcommand{\eea}{\end{eqnarray}}
\newcommand{\noi}{\noindent}
\newcommand{\non}{\nonumber}
\newcommand{\tstrut}{\rule[-1.8ex]{0ex}{4.5ex}}
\newcommand{\lsim}{\mathrel{\lower4pt \hbox{$\sim$}}
\hskip-12.5pt\raise1.6pt\hbox{$<$}\;}  
\newcommand{\gsim}{\mathrel{\lower4pt \hbox{$\sim$}}
\hskip-12.5pt \raise1.6pt \hbox{$>$}\;}
\begin{document}
\begin{titlepage}
\title{\begin{flushright}{\small \vspace{-.25in} BNL-HET-01/12}
\\ \end{flushright}
\Large \bf An absolute polarimeter for \\ high energy protons}
\author{\small \bf{N.~H.\ Buttimore}
\\ {\small \it School of Mathematics, University of Dublin, Trinity College,}
\\ {\small \it Dublin 2, Ireland}
\and
{\small \bf E. Leader}
\\ {\small \it Imperial College, Blackett Laboratory,}
\\ {\small \it Prince Consort Road, London SW7 2AZ, England}
\and
{\small \bf T.~L. Trueman}
\\ {\small \it Physics Department, Brookhaven National Laboratory,}
\\ {\small \it Upton, N.Y. 11973, U.S.A.}
}
\date{\small \today}
\maketitle
\normalsize
\begin{abstract}
 A study of the spin asymmetries for polarized elastic
 proton proton collisions in the electromagnetic hadronic
 interference (CNI) region
 of momentum transfer
 provides a method of self calibration of proton polarization.
 The method can be extended to non-identical spin half
 scattering so that, in principle, the polarization of a proton
 may be obtained through an analysis of its elastic collision
 with a different polarized particle, 
 \hel, 
 for instance.
 Sufficiently large CNI spin asymmetries
 provide enough information to facilitate the evaluation
 of nearly all the helicity amplitudes at small $t$ as well as the polarization
 of both initial
spin half fermions. Thus it can serve equally well as a polarimeter for
\hel.
\end{abstract}
\end{titlepage}
 
\section{Introduction
}
The RHIC collider in its $pp$ mode is 
a unique
machine.  It  opens 
a new frontier in the use of spin for the study
of hadronic physics. Using its polarized beams a whole new range of tests of
the Standard Model will become feasible, and much new information about  the
detailed partonic structure of the nucleon will emerge [1]. It  will
also be possible to answer intriguing questions concerning  the
relationship between $pp$ and $\bar pp$ total cross-sections and  real
parts of forward amplitudes, questions which are relevant to  attempts
to understand certain aspects of non-perturbative QCD [2]. The  entire
rich program relies upon an accurate determination of the  polarization
of the proton beams, hopefully to an accuracy of $\pm5\%$, a  matter
which is far from trivial.

Much effort has gone into attempts to  devise
a polarimeter with the required accuracy.   The attractiveness of using
a Coulomb Nuclear Interference  (CNI)
polarimeter to measure the polarization at very high energies  stems
from the belief, as folklore has it, that all hadronic  helicity-flip
amplitudes are negligible at high energies. In that case, and  ignoring the
real part of the non-flip amplitude, the analyzing power $A_N$ has  a
maximum at $t = t_p\equiv -8\pi\, \sqrt{3}\,\alpha /\sigma_{\rm tot}$
\cite{KL} and  its value at the peak is   
\be
A^{\rm max}_N = \frac{(-3t_p)^{1/2}}{2m}\cdot
\left(\frac{\kappa}{2} \right)  \label{eq1}
\ee   
\noi where $\kappa= \mu -1$ is the anomalous magnetic moment of the 
proton and $m$ its mass. (At
high energies $A^{\rm max}_N$ is between 4\% and 5\%.) Thus, in the  above
scenario, one has an almost perfectly calibrated  polarimeter.  
The problem is that if the helicity flip amplitudes are  non-negligible
then $\kappa/2$ in Eq.~(\ref{eq1}) has to be replaced  by   \be
\frac{\kappa}{2}\to\frac{\kappa}{2}-I_5  +\frac{\kappa}{2} I_2
\label{eq2}
\ee   \noi where $I_2$ and $I_5$ essentially measure the size of  the
imaginary parts of $\phi_2$ and $\phi_5$ relative to the imaginary  part
of the dominant non-flip amplitude $\frac12 (\phi_1+\phi_3)$.  The
precise definitions of $I_2$ and $I_5$ are given in terms of helicity
amplitudes in Eq.~(\ref{eq8}) to Eq.~(\ref{eq11}) below. The $\phi_i$'s are
defined in
\cite{GGMW}. Thus the efficacy of  the CNI Polarimeter rests upon being able
to demonstrate that
$I_5$  and
$I_2$ are negligible or knowing precise values for them.   In a very
comprehensive study of this issue
\cite{bigpaper} we concluded  that
while it was likely that $I_2$ can be neglected, it was not possible  to
exclude the possibility that $|I_5|\approx 15 \%$  implying
that the value of $A^{\rm max}_N$ can not be calculated to the  desired
accuracy of $\pm5\%$.  (A study of Coulomb interference with polarized protons, from
a rather different point of view, has been made by Jakob and Kroll \cite{J&K}.)

However, in the course of this study we discovered
the extraordinary
fact that $pp$ elastic scattering is self-calibrating, in the  sense
that a measurement of a sufficient number of spin-dependent  observables
at very small $t$ using a polarized beam and target, but without  \`a
priori knowledge of the magnitude of the polarizations, will  determine
not only the value of most of the helicity amplitudes at small  $t$,
but, remarkably, also the values of the beam and target  polarizations.  
This new approach to polarimetry was discussed briefly in \cite{bigpaper},
but without adequate analysis of the importance of certain  correction
terms.   

It is likely that some of the helicity amplitudes are very  small at
RHIC energies, in which case there are small correction terms that  were
not taken into account in \cite{bigpaper} which should be included.  
It is also possible that much larger spin asymmetries exist in  the
elastic collision of protons with some other spin 1/2 fermion  e.g.\
\hel, so that a more accurate determination of the  polarization
might be possible via such a reaction.   Finally, as compared with
\cite{bigpaper}, we have discovered a more  direct way
of evaluating the polarization from the measured asymmetries.  We
therefore present here a detailed analysis of the $pp$ case using  a
notation which allows an immediate generalization to the case  of
non-identical fermions, and we include correction terms which may  be
important at RHIC energies. We also present a simplified method  of
obtaining the polarization.   

The method we suggest involves the
taking of ratios of  experimentally
determined asymmetries, some of which may be very small.  Because so
little is known about the helicity amplitudes at high energy we  are
unable to study this question quantitatively. It will turn  out,
however, that there are several different ways of determining  the
polarization of the initial particles and one will have to  discover
pragmatically which of these provides the most accurate  determination.
In practice, one would use the values of all the measured  asymmetries to
extract a best fit to the polarizations.   

The method requires the use
of asymmetries with both  longitudinal and
transversely polarized beams. It will not succeed unless data  are
available with both configurations. Here we work only in order  $\alpha$
and so only amplitudes that are large compared to the next  order
correction can be determined from the formulas given below. (This  could
probably be improved upon if necessary: at present experiment  will
probably not be able to probe amplitudes below that size and so we  have
not pressed on in this direction.) We assume that the polarized  beams
have the same degree of polarization  in either configuration:  since
they are produced from the same initial configuration by rotation  this
is almost certainly true.   

Although it is likely that the two beams in a
$pp$ collider have the same polarization, our formulae will allow for
different polarizations $P$ and $P'$. This distinction will surely be
important for beams of different species.   

By now there are other methods that will be available for colliding proton
beam polarimetry and which may be simpler and more practical
than the one we discuss here. Nevertheless, we present the method because
it is elegant and it is interesting that $P$ can be determined in this way.
Furthermore, it may turn out that this is the best way for \hel \,
polarimetry,
which will very likely be needed eventually at RHIC. Section 5 presents an
alternative method for \hel\, polarimetry; it has some model dependence but will
almost certainly be practical for high energy $p$\,\hel \, collisions.

\section{Identical Fermions}  To see the principles of the method,
first consider the elastic collision of identical protons with mass $m$
and anomalous magnetic moment $\kappa$.   The usual five helicity amplitudes
$\Phi_j(s,t)$ \cite{GGMW} are  normalized
so that   \be
\sigma_{\rm tot}(s) =  \frac{4\pi}{\{s(s-4m^2)\}^{1/2}} {\rm Im\,}
(\Phi_1+\Phi_3)_{t=0}  \label{eq3}
\ee   \noi and   \be
\frac{d\sigma(s,t)}{dt} =\frac{2\pi}{s(s-4m^2)} \{  |\Phi_1|^2 + |\Phi_2|^2
+|\Phi_3|^2 + |\Phi_4|^2 +4|\Phi_5|^2\}  \label{eq4}
\ee   At the vary small values of $t$ we are interested in, the  interference
between the strong and electromagnetic forces is crucial, and  can be
taken into account by writing   \be
\Phi_j = \phi_j e^{-i\delta} + \phi^{em}_j  \label{eq5}
\ee   
\noi where $\phi_i$ are the hadronic amplitudes and $\delta$  is the
Coulomb phase which was shown in \cite{BGL} to be the same for all  helicity
amplitudes. Its role will be discussed in Section~4. 
In \eq{eq5}  the
$\phi^{em}_i$ are one photon exchange amplitudes and are real.   We are, of
course, primarily interested in learning about the  hadronic
amplitudes $\phi_i$ for which we shall write expressions valid for  very
small $t$.   We define   \be
\phi_\pm \equiv \frac12 (\phi_1\pm\phi_3)  \label{eq6}
\ee   \noi so that at high energies Im$\phi_+(s,0)$ is large and  positive and
should be the dominant amplitude.   We shall put    \be
\phi_+(s,t) = {\rm Im\,} \phi_+ \,\lp\rho \, e^{\bar{B}t/2}  +  i
e^{Bt/2}\rp
\label{eq7}
\ee   
\noi where Im$\phi_+$ is short for Im$\phi_+(s,0)$, $\rho$ is  the usual
ratio of real to imaginary parts of the forward spin averaged  amplitude
and $B$ is the slope of $d\sigma/dt$.  Here we allow the possibility that the real
and imaginary parts have somewhat different slopes. We also
introduce scaled amplitudes as follows:  
\bea
\lp R_2  e^{\bar{B}_2t/2}+ iI_2  e^{B_2t/2}\rp & = & 
\frac{\phi_2(s,t)}{2\,{\rm Im\,}\phi_+}
\label{eq8} \\[1ex]
\lp R_-e^{\bar{B}_-t/2} + iI_-  e^{B_-t/2}\rp & = & \frac{\phi_-(s,t)}{{\rm
Im\,}\phi_+}
\label{eq9}  \\[1ex]
\lp R_5e^{\bar{B}_5t/2} + iI_5 e^{B_5t/2}\rp& = &
\left(\frac{m}{\sqrt{-t}}  \right)
\frac{\phi_5(s,t)}{{\rm Im\,}\phi_+} \label{eq10}  \\[1ex]
\lp R_4e^{\bar{B}_4t/2} + iI_4  e^{B_4t/2}\rp & = &
\left(\frac{m^2}{-t}\right)\frac{\phi_4 (s,t)}{{\rm Im\,}\phi_+}
\label{eq11}
\eea  
Note the factor of 2 in \eq{eq8}, which is  introduced for
later convenience. Note too that we have taken out explicit  kinematic
factors of $\sqrt{-t}$ and $t$ from $\phi_5$ and $\phi_4$  respectively
since for the strong interaction amplitudes  $\phi_5\propto\sqrt{-t}$ and
$\phi_4\propto t$ as $t\to0$.     The
$R_j$ and
$I_j$  may vary with $s$.   

We shall need to consider various experimentally 
determined asymmetries, $PA_N\frac{d\sigma}{dt}$,  $P^\prime
A_N\frac{d\sigma}{dt}$, $PP^\prime A_{NN}\frac{d\sigma}{dt}$,  $PP^\prime
A_{LL} \frac{d\sigma}{dt}$, $PP^\prime A_{SS}\frac{d\sigma}{dt}$  and
$PP^\prime A_{SL}\frac{d\sigma}{dt}$ \cite{Bourrely:1980mr}.   Those contain
singular terms as
$t\to0$ arising
from the  interference between one photon exchange and hadronic amplitudes.
To order
$\alpha$ the asymmetries $A_{NN}\f{d\sigma}{dt}$,
$A_{LL}\frac{d\sigma}{dt}$
and $A_{SS}\frac{d\sigma}{dt}$ are singular as  $1/t$ while
$A_N\frac{d\sigma}{dt}$ and $A_{SL}\frac{d\sigma}{dt}$ vary as
$1/\sqrt{-t}$.    We shall also require the total cross-section differences
$PP^\prime\Delta
\sigma_T$ and $PP^\prime\Delta\sigma_L$. Keeping therefore  the most
singular terms as $t\to0$ and working to order $\alpha$, which  implies
ignoring the Coulomb phase $\delta$ in (\ref{eq5}) (the re-insertion  of
$\delta$ will be discussed in Section 3), we may write the  following
expressions for the various experimental observables:   
\bea
\Delta_T & \equiv -\frac12 
PP^\prime\displaystyle \frac{\Delta\sigma_T}{\sigma_{\rm tot}} & = PP^\prime
I_2 
\label{eq12} \\
\Delta_L & \equiv \phantom{-}\frac12 
PP^\prime\displaystyle\frac{\Delta\sigma_L}{\sigma_{\rm tot}} & = PP^\prime
I_-  \label{eq13}
\eea  
\noi For the differential cross section, with slope parameter  $B$
for
proton proton collisions, which has an expansion in powers of  $t$
beginning with the photon pole and interference terms we write the  form
\cite{Amaldi}  
\be
{\cal I}_0 \equiv \left(\frac{t}{e^{Bt}\sigma_{\rm  tot}}\right)
\frac{d\sigma}{dt} = \frac{4\pi}{\sigma_{\rm tot}}\f{\alpha^2}{t} + 
\alpha\, a_0+ \frac{\sigma_{\rm tot}}{8\pi} b_0\,t .\label{eq14}
\ee  
Here, and in the expressions to follow, it is assumed that  coefficients
like $a_j$ and $b_j$ (see Table~\ref{tabone}) are  essentially
$t$-independent in the very small range of $t$ under  consideration. (Note
that we have changed the signs of some of the definitions of these
quantities from \cite{bigpaper} for clarity. Also, there was a sign error
in the table entry for $A_N$ in \cite{bigpaper} which has been corrected
here.) This key assumption can be tested experimentally by confirming  that
the observables really do follow the
$t$-dependence given.  
\begin{table}[htb]
\begin{center}
\caption{Expressions  for the coefficients $a_j$ and $b^{(0)}_j$
relevant to the measured  observables for $pp$ elastic scattering
\label{tabone}}
\vspace{.1in}
\begin{tabular}{|r|l|l|}
\hline
\tstrut Observable & $a_j$ & $b^{(0)}_j$  \\
\hline
\tstrut ${\cal I}_0$ & $\rho$ & $\frac12  (1+\rho^2+R^2_-+I^2_-) + R^2_2
+I^2_2$
\\
\hline
\tstrut$PP^\prime  A_{NN}
{\cal I}_0 
$ & $PP^\prime R_2$ & $PP^\prime [R_2
(\rho+R_-)  +I_2 (1+I_-)]$ \\
\hline
\tstrut$PP^\prime A_{LL}%
{\cal I}_0
$ &  $PP^\prime R_-$ & $PP^\prime [\rho R_-
+I_- +R^2_2 +I^2_2]$  \\
\hline
\tstrut$PP^\prime A_{SL} {\cal I}_1$ & $-PP^\prime  \frac{\kappa}{2}
(R_2+R_-)$ & $-PP^\prime [R_5(R_2+R_-) + I_5 (I_2+I_-)]$  \\
\hline
\tstrut$P A_N {\cal I}_1$ &  $\ph{P'}P[\frac{\kappa}{2}(1+I_2)-I_5  ]$ & $\ph{P'}
P[-I_5 (\rho+R_2) +R_5 (1+I_2)]$  \\
\hline
\tstrut$P^\prime A_N {\cal I}_1$ & $\ph{P'}P^\prime [\frac{\kappa}{2}
(1+I_2)-I_5]$ & $\ph{P'}P^\prime[-I_5(\rho+R_2) +R_5 (1+I_2)]$  \\ 

\hline
\end{tabular}
\end{center}
\end{table}  
The quantities $b^{(0)}_j$ are calculated assuming that all the hadronic
slopes are equal and equal to the electromagnetic slopes as well. This
will be corrected for below. For asymmetries with initial protons polarized
parallel or  anti-parallel along the same axis, $N$ (normal to the
scattering plane) or 
$L$ (longitudinal), we have    \bea
PP^\prime A_{NN} {\cal I}_0 & = & \alpha  a_{NN} + \frac{\sigma_{\rm
tot}}{8\pi} b_{NN} t+\cdots \label{eq15}  \\
PP^\prime A_{LL} {\cal I}_0 & = & \alpha a_{LL}  +\frac{\sigma_{\rm
tot}}{8\pi} b_{LL} t+\cdots  \label{eq16}
\eea   \noi Spin observables with both initial protons polarized  either way
along perpendicular axes have a similar power series  expansion   
\be
PP^\prime A_{SL} {\cal I}_1 = \alpha \, a_{SL}  +\frac{\sigma_{\rm
tot}}{8\pi}\, b_{SL} \,t \label{eq17}
\ee  
\noi where again the coefficients $a_{SL}$, $b_{SL}$ are given  in
Table~\ref{tabone} and we have defined
\be
{\cal I}_1 \equiv \frac{m\sqrt{-t}}{e^{Bt}\sigma_{\rm  tot}} \,
\frac{d\sigma}{dt} \label{eq18}
\ee
\noi Of course for $pp$ scattering  $A_{LS}=A_{SL}$.   Note that to
the accuracy of our approximations  $A_{NN}=A_{SS}$, so
the latter is not discussed separately. However a  measurement of
$A_{SS}$ could be used as a check on the adequacy of  our
approximations.   For spin asymmetries with only one of the initial protons
polarized we have  
\bea
P A_N {\cal I}_1 & = & \alpha \,a_N 
+  
\frac{\sigma_{\rm tot}}{8\pi}\, b_N\,t+\cdots, \label{eq19}
\\[1ex]  
P^\prime A_N  {\cal I}_1 & = & \alpha \,a_N\,' 
+  
\frac{\sigma_{\rm tot}}{8\pi} \, b_N\,'\,t+\cdots .\label{eq20}
\eea  
\noi We write the last pair in this rather pedantic way to  emphasize
the fact that single spin asymmetries may be measured for either  beam
and, although the analyzing powers are the same, the polarizations  $P$
and $P^\prime$ may be different. For the comparable situation  discussed
in the Section 4, the analyzing powers will be different too,  in
general. The expressions for the measurable coefficients, $a_j$, $b_j$,  taken
from \cite{bigpaper}, are given in Tables~\ref{tabone} and \ref{tabtwo} in 
terms of the hadronic amplitudes $R_j$, $I_j$ and the anomalous  magnetic
moment $\kappa$. Each $b_j$ is written in the form   \be
b_j = b^{(0)}_j +\frac{8\pi\alpha}{\sigma_{\rm tot}} \, \Delta b_j
\label{eq21}
\ee  
\noi with the $b^{(0)}_j$ given in Table~\ref{tabone} and the  $\Delta
b_j$ in Table~\ref{tabtwo}. Note that $b^{(0)}_{SS} = b^{(0)}_{NN}$,  so
it is not listed.  We have kept only the  largest
correction term for $b_N$.  
\begin{table}[htb]
\begin{center}
\caption{Correction  terms $\Delta b_j$ in $b_j=b^{{0}}_j +
\frac{8\pi\alpha}{\sigma_{\rm tot}} \Delta  b_j$ \label{tabtwo}}
\vspace{.1in}
\begin{tabular}{|r|l|}
\hline
\tstrut Observable &  \qquad\qquad $\Delta b_j$ \\
\hline
\tstrut$PP^\prime A_{NN}{\cal I}_0 $  & $ PP^\prime \left[
\frac{\kappa^2}{4m^2} \rho +R_2 \,(2\beta_1+\bar{B}_2/2-B) -\frac{\kappa}{m}
R_5 +\frac{R_4}{2m^2}\right]$  \\
\hline
\tstrut$PP^\prime  A_{SS}{\cal I}_0$ & $PP^\prime \left[
\frac{\kappa^2}{4m^2} R_- +R_2 \, (2\beta_1+\bar{B}_2/2-B) -\frac{R_4}{2m^2}
\right]$  \\
\hline
\tstrut$PP^\prime A_{LL}{\cal I}_0$ & $PP^\prime \left[R_-  (2\beta_1+
\bar{B}_-/2-B) +\frac{\kappa^2}{4m^2} R_2\right]$  \\
\hline
\tstrut$PP^\prime A_{SL} {\cal I}_1$ & $-PP^\prime \, \frac{\kappa}{2}
\,\left[R_2\, (\beta_1+\beta_2+\bar{B}_2/2-B) +R_- (\beta_1+\beta_2+
\bar{B}_-/2-B) -\frac{R_4}{2m^2} \right]$ \\
\hline
\tstrut$PA_N{\cal I}_1$ & $\ph{.}P \frac{\kappa}{2}\, (\beta_1+\beta_2 -B/2)$  \\
\hline
\end{tabular}
\end{center}
\end{table}  
In these expressions we have not included the Coulomb  phase $\delta$.
We shall explain in Section 3 the consequences of including  it. Note
that $\phi_4$ only occurs in the correction terms $\Delta  b_j$.  
In Table~\ref{tabtwo} the parameters $\beta_1$, $\beta_2$ have  the
following significance. In order to take into account  the
$t$-dependence of the electromagnetic form factors of the proton  we
parametrize the Dirac and Pauli form factors, for very small $t$, in the form
\be
F_1(t) = e^{\beta_1t} \,, \qquad F_2(t) = e^{\beta_2t} \, \label{eq22}
\ee
\noi where, it turns out that with \( \Lambda^2 = 0.71 \m{ GeV}^{-2}
\) \cite{emforms},
\bea
\beta_1 & = & 2/\Lambda^2 - \kappa/(2m)^2 \, = \, 2.31 \m{ GeV}^{-2}, \nonumber
\\[1ex]
\beta_2 & = & 2/\Lambda^2 + 1/(2m)^2 \, = \, 3.10 \m{ GeV}^{-2} \,.
\label{eq23}
\eea  
Strictly, in order to use our method, one must have
information about the  slopes
$\bar{B}_2$ and
$\bar{B}_-$ , which seem very hard to come by, or else to neglect these
corrections. Since they always appear in the form of a difference from $B$
times one of the small amplitudes it seems reasonable to neglect them in what
follows. (Of course the correction from the difference between the electromagnetic
slopes and $B/2$ is known and that correction can be applied.)

\section{The Method Applied to Identical Particles
}
We assume that $\sigma_{\rm tot}$, $d\sigma/dt$, $\rho$,  $\Delta_L$ and
$\Delta_T$ (see \eq{eq12} and \eq{eq13}) are  measured, and
that some of the coefficients in \eq{eq15} to \eq{eq20}, in
particular $a_{LL}$, $b_{LL}$ and  $a_{NN}$ have been determined.    these
provide values for the product of the unknown  polarizations with
the real parts of the non-zero forward  amplitudes:   \be
\rho=a_o \qquad PP^\prime R_2=a_{NN} \qquad PP^\prime  R_-=a_{LL}
\label{eq24}
\ee   Substituting these results into the expression for $b_{LL}$  in
Tables~\ref{tabone} and \ref{tabtwo} one has:   \bea
b_{LL} & = & PP^\prime \left[ \rho  R_-+I_-+R^2_2+I^2_2
+\frac{2\pi\alpha\kappa^2}{m^2\,\sigma_{\rm tot}} R_2  \right] \non \\
& = & \rho a_{LL} +\Delta_L + (a^2_{NN}  +\Delta^2_T)/PP^\prime
+\frac{2\pi \alpha\kappa^2}{m^2\, \sigma_{\rm tot}}  a_{NN} \label{eq25}
\eea   
\noi from which one obtains an expression for the  polarization   
\be
PP^\prime =\lp a^2_{NN}+\Delta^2_T \rp  / \lp b_{LL} -\rho  a_{LL}-\Delta_L
-\displaystyle\frac{2\pi\alpha\kappa^2}{m^2\,\sigma_{\rm tot}} a_{NN}\rp . 
\label{eq26}
\ee  

There is an important question which we will try to address.  How
reliable is \eq{eq26} if the helicity amplitudes are small?  Although
we cannot predict the magnitude of the helicity amplitudes at  high
energies, consideration of quantum number exchange \cite{bigpaper} suggests 
that
$|\phi_-|$ will be the smallest of the above hadronic amplitudes,  and
we expect $|R_-|$, $|I_-|<<1\%$ \cite{Grosnick}. For the method to
work we need
$|\phi_2|$ to be  considerably larger, albeit still small. How small? In
order to use \eq{eq26} at all it is clear that one must have the
uncertainties $\Delta (b_{LL}) \ll b_{LL}$ and $\mbox{max of }  
(\Delta(\Delta_L )\, \mbox{and } \Delta(\Delta_T)) \ll \mbox{either }
a_{NN}\; \mbox{or } \Delta_T$. If this is false then our whole approach would break
down even  if
$R_5$ and $I_5$ are large. This is because the reaction then  becomes
analogous to spin 1/2\,-\,spin 0 scattering, for which we certainly  cannot
determine the amplitudes without knowing $P$.  
The requirement that  a relative
error in $P P'$ less than $10\%$ is obtained in this way is more
demanding; for then using \eq{eq26} we will need (assuming for illustration that all
the needed observables $a_i, b_i, \Delta_i$ have the same relative error
$\epsilon$ and adding the errors in quadrature) 
$\epsilon < 0.1/\sqrt{5}$. Thus if  $b_{LL}$ turns out to be less than 0.1,
the error will need to be less than about 0.005 which may be doable, but
much smaller will probably not be accessible.

If the first condition is satisfied, but the second is not, so that 
$PP^\prime$ determined via \eq{eq26} has a large uncertainty, there 
is a more reliable though indirect route to determine the  polarization,
which involves first determining $I_2$ and $I_5$ and then using  them to
calculate the analyzing power $a_N$ (see Table~\ref{tabone}).  Since
they will provide only a small correction to the dominant  term
$\kappa/2$, it could well be that the errors in their determination  are
unimportant in $A_N$.   

To achieve this,    we assume that we
have a rough determination  of
$PP^\prime$ from (\ref{eq26}) and use it to provide an improved  result.  
Let us define the combination of observables in (\ref{eq26})  as   \be
O_1 \equiv \lp b_{LL} -a_{LL} -\Delta_L  -\displaystyle\frac{2\pi\alpha\kappa^2
a_{NN}}{m^2\sigma_{\rm tot}}\rp / \lp a^2_{NN}+\Delta^2_T \rp  \label{eq28}
\ee   \noi Then from (\ref{eq24}) we have an estimate of the real  parts:   \be
R_2 = O_1 a_{NN} \qquad R_- =O_1a_{LL}  \label{eq29}
\ee   \noi Also from (\ref{eq12}) and (\ref{eq13}) we have an  estimate of the
imaginary parts:   \be
I_2=O_1\Delta_T \qquad I_- = O_1\Delta_L  \label{eq30}
\ee   We now proceed to obtain an estimate of $R_5$ and $I_5$. Using  $a_{SL}$
and the uncorrected expression for $b_{SL}$ from  Table~\ref{tabone}
yields   \be
R_5 + \left(\frac{I_2+I_-}{R_2+R_-} \right) I_5 =  \frac{\kappa}{2}\,
\frac{b_{SL}}{a_{SL}} \label{eq31}
\ee   \noi which, via (\ref{eq12}), (\ref{eq13}) and  Table~\ref{tabone}
becomes   \be
R_5 + \left( \frac{\Delta_T+\Delta_L}{a_{NN}+a_{LL}}  \right) I_5 =
\frac{\kappa}{2}\, \frac{b_{SL}}{a_{SL}}  \label{eq32}
\ee   The asymmetry $A_N$ provides a second relation between $R_5$  and $I_5$.
From Tables~\ref{tabone} and \ref{tabtwo} we have   \be
\frac{b_N}{a_N} = \frac{I_5\,(\rho+R_2)  -R_5\,(1+I_2)}{I_5
-\frac{\kappa}{2} (1+I_2)} +  \frac{8\pi\alpha
(\beta_1+\beta_2-B/2)}{\sigma_{\rm tot}}  \label{eq33}
\ee   \noi where a small term has been neglected in the denominator  of the
connection term. This can be rewritten as   \be
R_5 + \left( \frac{c_N -\rho -R_2}{1+I_2}\right) I_5  =
\frac{\kappa}{2} c_N \label{eq34}
\ee   \noi where   \be
c_N \equiv \frac{b_N}{a_N} -  \frac{8\pi\alpha\,
(\beta_1+\beta_2-B/2)}{\sigma_{\rm tot}}  \label{eq35}
\ee   \noi and the known last term in (\ref{eq35}) may turn out to  be
negligible.   Solving (\ref{eq32}) and (\ref{eq36}) for $I_5$ and $R_5$,  using
(\ref{eq30}) we obtain   \be
I_5 = \frac{\displaystyle \frac{\kappa}{2} (\frac{b_{SL}}{a_{SL}}  -c_N)}
{\left(\displaystyle\frac{\Delta_T+\Delta_L}{a_{NN}+a_{LL}} \right) - 
\left(\displaystyle\frac{c_N -\rho-a_{NN}O_1}{1+O_1\Delta_T}\right)} 
\label{eq36}
\ee   \noi and   \be
R_5 = \frac{\kappa}{2} \left\{ \frac{b_{SL}}{a_{SL}}  -
\frac{\displaystyle\frac{b_{SL}}{a_{SL}}  -c_N}{1-
\displaystyle\frac{a_{NN}+a_{LL}}{\Delta_T+\Delta_L}\,\displaystyle\frac{c_N
-\rho-a_{NN}O_1}{1+\Delta_TO_1}}  \right\}. \label{eq37}
\ee   

Now if it turns out that $I_5$ is very small, say  $\lsim1\%$, then
the neglect of the correction term in the expression for  $b_{SL}$
renders our estimate inaccurate. However $I_5$ is then so  small
compared to $\kappa/2$ in the expression for $a_N$ that is  irrelevant
and should be neglected. If, on the other hand it is ``large''  i.e.\
$\approx 10\%$ then the neglected correction term in $b_{SL}$  is
negligible and the estimate for $I_5$ (and $R_5$) can be used  in
computing the hadronic corrections to $A_N$.   As stressed earlier,
since $I_5$ and $I_2$ are small  corrections
compared to $-\kappa/2$ in $a_N$, the errors on them, even  if
relatively large ---and they may be, given the large number of
measured quantities that go into \eq{eq36} and \eq{eq37}--- should not be
important in computing the value  of
$a_N$.   Once $A_N$ is known the polarization can be measured in the
normal way.

Up to the present
we have ignored the Coulomb phase $\delta$  which appears in (\ref{eq5}). We
now consider the effect of including it.  It will turn out that $\delta$ has
{\it no\/} effect upon  the determination of the polarization given in
Section~3. Its  only role will be in the evaluation of the hadronic
amplitudes  themselves.    It is given by \cite{bethe, cahn, KT1}
\be
\delta = -\alpha [\ln {(|t| (B/2 + 4/\Lambda^2))} +\gamma]  \label{eq39}
\ee  
\noi where $B$ is the logarithmic slope of the $pp$  $d\sigma/dt$ at
$t=0$ , $\Lambda^2 = 0.71 \, {\rm GeV}^2$ and ${\gamma}$ is Euler's constant
$\gamma=0.5772$. Experimentally
$B\approx 13$ GeV$^{-2}$ at high energies and increases  slowly through
the RHIC energy range.   The phase $\delta$ is very small except at
extremely tiny  values of
$|t|$, so that we can take   \be
\phi_j e^{-i\delta} \approx ({\rm Re} \phi_j +\delta \, {\rm Im\,} \phi_j)
+i {\rm Im\,}\phi_j \label{eq40}
\ee  
Thus to take into account $\delta$ one must make the  replacements   \be
R_j\to R_j+ I_j \,\delta \label{eq41}
\ee   \noi and   \be
\rho \to \rho+\delta \label{eq42}
\ee   \noi throughout Sections 2 and 3.  
Note that, strictly speaking, $\delta$ is a function of $t$.  However it
seems hopeless to try to detect the $\ln{|t|}$ behavior in  \eq{eq39},
so we suggest that \eq{eq41} and \eq{eq42} should be used  with an
effective $t$-independent $\bar\delta$, equal to the mean value  of
$\delta(t)$ for the region of $t$ covered experimentally.   It is important
to note that the substitutions \eq{eq41} and
\eq{eq42} do {\it not\/} affect the result \eq{eq26}  for
$PP^\prime$, nor the result for $A_N$ calculated on the basis of  the
Table~\ref{tabone} and \ref{tabtwo} formula, in which the  results
\eq{eq36}, \eq{eq37}, \eq{eq29} and \eq{eq24} are  used.  
The phase $\delta$ only plays a role in the determination of  the
hadronic amplitudes {\it per se\/}; i.e.\ any amplitude derived  via
\eq{eq24}, \eq{eq29}, \eq{eq36} and \eq{eq37} must  be
reinterpreted on the basis of \eq{eq41} and \eq{eq42}.              
\section{Nonidentical spin half fermions
}
 The above analysis generalizes to the
 case of the collision of non-identical spin half particles,
 protons colliding with \hel \, in the CNI region, for example.
 It may be that the self calibration of proton polarization through
 the use of proton proton CNI collisions is difficult at particular
 energies because of unsuitably small spin asymmetries.
 In such cases it is, in principle, possible to evaluate the proton
 polarization by a study of the non-identical collision process.

 Suppose that the second distinct fermion of mass $m'$ and
 polarization $P'$ has  charge $Ze$, and anomalous magnetic moment
$\kap'$, given in units of proton magnetons, as is usual for nuclei.
Real and imaginary scaled amplitudes are introduced as before
 that now involve the centre of mass momentum $k$ defined by
\(
        4 k^2 s  =  [s - (m' - m)^2] [s - (m' + m)^2].
\)
 A distinct helicity flip amplitude for the second fermion 
\ba
\lp R_6 e^{\bar{B}_6t/2} + iI_6  e^{B_6t/2}\rp
& = &
 \, \left(\frac{m}{\sqrt{-t}}  \right)
\frac{\phi_6(s,t)}{{\rm Im\,}\phi_+} \label{eq:phi6}
\ea
 appears in non-identical elastic collisions. 

 Using only initial state polarization, with one or
both beams polarized,
one can measure seven spin dependent asymmetries. We follow the
notation of
\cite{BGL}; notice that $\phi_5$ describes the proton spin-flip while $\phi_6$
describes the \hel \, spin-flip, while $A_N$ is the proton analyzing power and $A_N'$
is the \hel \, analyzing power.
\begin{eqnarray} \label{eq:asymdef}
       \frac{d\sigma}{dt} & = & \frac{ \pi}{2k^2s}
\left\{
\,	|\phi_1|^2 + |\phi_2|^2 + |\phi_3|^2 + |\phi_4|^2
\non	+
	2 |\phi_5|^2 + 2 |\phi_6|^2
\right\}
\gp
	A_{N} \frac{d\sigma}{dt} & = & -\frac{\pi}{k^2s} \im
\lb
	\lp \phi_1 + \phi_3 \rp \phi_5^\ast
\non	-
	\lp \phi_2 - \phi_4 \rp \phi_6^\ast
\rb
\gp
	A_{NN} \frac{d\sigma}{dt} & = &	\frac{ \pi}{k^2s} \re
\lb
\non	\phi_1 \phi_2^* - \phi_3 \phi_4^* - 2\phi_5\phi_6^\ast
\rb
\gp
	A_{N}'  \frac{d\sigma}{dt} & = & \frac{ \pi}{k^2s} \im
\lb
	\lp \phi_1 + \phi_3 \rp \phi_6^\ast
\non	-
	\lp \phi_2 - \phi_4 \rp \phi_5^\ast
\rb
\gp
	A_{SS} \frac{d\sigma}{dt} & = &	\frac{ \pi}{k^2s} \re
\lb
	\phi_1 \phi_2^* + \phi_3 \phi_4^* \non
\rb
\gp
	A_{SL} \frac{d\sigma}{dt} & = & \frac{ \pi}{k^2s} \re
\lb
	\lp \phi_2 + \phi_4 \rp \phi_5^*
	-
	\lp \phi_1 - \phi_3 \rp \phi_6^* \non
\rb
\gp
	A_{LS} \frac{d\sigma}{dt} & = & \f{ \pi}{k^2s} \re
\lb
	\lp \phi_1 - \phi_3 \rp \phi_5^*
	-
	\lp \phi_2 + \phi_4 \rp \phi_6^* \non
\rb
\gp
	A_{LL} \frac{d\sigma}{dt} & = & \f{ \pi}{2k^2s}
\left\{
\,	|\phi_1|^2 - |\phi_3|^2 + |\phi_2|^2 - |\phi_4|^2 \,
\right\} \,.
\end{eqnarray}

 For scattering of a proton on a spin-$1/2$ fermion of
charge $Z$ and anomalous magnetic moment $\kappa'$ the electromagnetic
amplitudes are approximately, for large $s$ and small $t$,
\bea \label{eq:1 photon ex.}
	\aml{1}{em} & = & \ph{-} \aml{3}{em}
\,\,\,	=
\,\,\,	\f{\alpha s Z}{t} \, F_1 \, F_1^\prime \non
\gp
	\aml{2}{em} & = & -\aml{4}{em} \non
\,\,\,	=
\,\,\,	\f{\alpha s \kappa \kappa'}{4 m^2} \, F_2 \, F_2^\prime
\gp
\non    \aml{5}{em} & = & -\f{\alpha s \kappa Z}{2 m\s{-t}}
\,	F_1' \, F_2
\gp
  \aml{6}{em} & = & \ph{-} \f{\alpha s \kappa '}{2 m\s{-t}}
\,	F_1 \, F_2' \,
\eea
        where $\mu_p = \kap + 1$ is the proton's magnetic moment,
        and $m$ its mass.
 We recall that the proton electromagnetic form factors
 are denoted $F_1(t)$ and $F_2(t)$ while the Dirac and Pauli
 form factors of \hel\ we will describe by
 $F_1'(t)$ and $F_2'(t)$. All of the form factors are normalized to 1 at $t=0$.
 The form factor $F_1'$ appears multiplied by
 the charge of \hel ,\,$Z = 2$, but $F_2'$ does not. 
Rather it appears with the factor $\kappa'$; note that the magnetic moments are
given in units of nuclear magnetons, as is normal for nuclei, and so the proton mass
$m$ appears in the denominator of the expression for $\aml{6}{em}$ in \eq{eq:1 photon
ex.} and in the numerator of \eq{eq:phi6}.

 ${\cal I}_0$ and ${\cal I}_1$ are defined as before, and
\be
{\cal I}_0
\left(\frac{t}{e^{Bt}\sigma_{\rm  tot}}\right) \frac{d\sigma}{dt}
\,\, =
\,\, \frac{4\pi}{\sigma_{\rm tot}}\f{ Z^2 \alpha^2}{t}
 +
 \alpha \, a_0
 +
 \frac{\sigma_{\rm tot}}{8\pi} \, b_0\,t.
 \ee
 
\begin{table}
\caption{
        The left column shows the reaction parameter, an asymmetry
        in all cases excepting the first, the unpolarized cross section.
        The second and third columns indicate the related expressions
        for the coefficients $ \cfa{I} $ and $ \cfb{I} $
	.}
\label{tab.nonident}
\begin{center}
%
\begin{tabular}{|r|l|l|}        \hline
        Observable &  $ \cfa{I} $ & $ \cfb{I} $
        \rule[-1.3ex]{0ex}{4.0ex}
\gp
\hline  \rule{0ex}{4.0ex}  
$       {\cal I}_0                $&$ \qquad Z \rho
$&
$        \half \lp 1 + \rho^2 + \rc{-}^2 + \ic{-}^2 \rp
         +
         \ra_2^2 + \ia_2^2$
\gp
\hline		\rule{0ex}{4.0ex}
$       P P' A_{NN}{\cal I}_0 $&$ P P' \, Z \ra_2
$&
$       P P' \lb \ra_2(\rho + \rc{-}) + \ia_2(1 +  \ic{-}) \rb$
\gp
\hline	\rule{0ex}{4.0ex}
$       P P' A_{LL} {\cal I}_0 $&$ P P' \, Z \ra_-
$&
$       P P' \lp \rho\rc{-} + \ic{-} + \ra_2{}^2 + \ia_2{}^2 \rp
$
\gp
\hline \rule{0ex}{4.0ex}  
$      P P' A_{SL} {\cal I}_1
$&
$	-P P' \lp \lama Z \ra_2 + \lama' \rc{-} \rp
$&
$       P P' \lp -\ra_2 \ra_5 -\ia_2 \ia_5 + \rc{-}\ra_6 + \ic{-}\ia_6 \rp$
\gp
\hline	\rule{0ex}{4.0ex}
$         P  A_{N} {\cal I}_1
$&
$	 \ph{P'} P \lp \lama Z + \lama'\ia_2 - Z \ia_5 \rp
$&
$\ph{P'} P \lp -\rho\ia_5 + \ra_5 + \ra_2 \ia_6 - \ia_2 \ra_6 \rp$
\gp
\hline	\rule{0ex}{4.0ex}
$      P P' A_{LS} {\cal I}_1
$&
$	-P P' \lp \lama' \ra_2 + \lama Z \rc{-} \rp
$&
$       P P' \lp \ra_2 \ra_6 + \ia_2 \ia_6 - \rc{-}\ra_5 - \ic{-}\ia_5 \rp$
\gp
\hline	\rule{0ex}{4.0ex}
$   \ph{P}P' A_{N}' {\cal I}_1
$&
$	\ph{P'} P'\lp \lama' + \lama Z \ia_2 + Z \ia_6 \rp
$&
$   \ph{P}P' \lp \rho\ia_6 - \ra_6 - \ra_2 \ia_5 + \ia_2 \ra_5 \rp$
\gp
\hline
\end{tabular}
\gp
\end{center}
\end{table}

In addition to $A_{N}\,'$, which is truly independent of $A_{N}$ here, one
new non-trivial spin observable for the
 case of distinct fermions involves $A_{LS}$
 in addition to $A_{SL}$, with an additional equations corresponding to
\eq{eq17}
 \be
PP^\prime A_{LS} {\cal I}_1 = \alpha a_{LS}  +\frac{\sigma_{\rm
tot}}{8\pi} b_{LS}\, t \label{eq46}
\ee 
 The analysis proceeds as in the identical particle case up to
 the point just before consideration of the evaluation of $\ra_5$.
 To determine values of $\ra_5$ and $\ra_6$, now, we observe from the
 expressions for \A{SL} and \A{LS} in Table\ \ref{tab.nonident} that
\ba
        \cfa{SL} \pm \cfa{LS}
&       =
&       -P P' \half \lp \kap Z \pm \kap' \rp \lp \ra_2 \pm \rc{-} \rp \non
\gp
        \cfb{SL}\pm \cfb{LS} 
&       =
&      -P P' \lp \ra_5 \mp \ra_6 \rp  \lp \ra_2 \pm \rc{-} \rp
        -
       P P'  \lp \ia_5 \mp \ia_6 \rp \lp \ia_2 \pm \ic{-} \rp. \label{eq.diff}
\ea
      
        Upon division, we are led to two linear relations among the
        four quantities
        $\ra_5$, $\ra_6$, $\ia_5$, and $\ia_6$
        so that from \eq{eq.diff}
\ba
        \lp \ra_5 \mp \ra_6 \rp + \lp \ia_5 \mp \ia_6 \rp
        \f{ \ia_2 \pm \ic{-} }{ \ra_2 \pm \rc{-} }
&       =
&      \half \lp \kap Z \pm \kap' \rp            \label{eq.56two}
        \f{ \cfb{SL} \pm \cfb{LS} }{ \cfa{SL} \pm \cfa{LS} } \, .
\ea
        Two further linear relations among
        $\ra_5$, $\ra_6$, $\ia_5$, and $\ia_6$
        may be obtained from consideration of \A{N} and \A{N}$'$.

\begin{equation}
        \ra_5 - \ia_2 \ra_6      
         - \rho  \ia_5 +\ra_2 \ia_6
      =
      \half  \lp \kap Z + \kap'\ia_2 -2 Z I_5\rp \,
\f{\cfb{N}}{\cfa{N}},\label{eq.56three}
        \end{equation}
\be    
   \ia_2 \ra_5 - \ra_6      
     -\ra_2 \ia_5+\rho I_6 
       =
     \half \lp \kap' + \kap Z \ia_2 +2 Z \ia_6\rp \,
\f{\cfb{N}'}{\cfa{N}'} \label{eq.56four}
.
\ee
        From the four linear relations Eq.\ (\ref{eq.56two}--\ref{eq.56four})
        values of $\ra_5$, $\ra_6$, $\ia_5$, and $\ia_6$ may be found
        in general, just as in \eq{eq36} and \eq{eq37}.  A glance at $P A_{_{N}}$
and
$P' A_{_{N}}'$ in
        Table\ \ref{tab.nonident} reveals that the equations
\ba
        \cfa{N}                                 \label{eq.aN}
&       =
&       P \lp \lama Z + \lama'\ia_2 - Z \ia_5 \rp
\gp
        \cfa{N}\,' & = &  P'\lp \lama' + \lama Z \ia_2 - Z \ia_6 \rp
\ea
        provide estimates of the individual polarizations, $P$ and $P'$,
        of the two distinct fermions participating in the elastic collision.
        All the ingredients for successful polarimetry are to hand.
        If the various coefficients $\cfa{I}$ and $\cfb{I}$ involved in
        the evaluation of the scaled amplitudes are sufficiently
        non-zero within error, then Eq.\ (\ref{eq.aN}) indicates the
        extent of the asymmetry maximum to be expected for polarized
        protons. The corrections and error discussion go through just as in
Section 3.
\section{An alternative method for ${}^3$He polarimetry}
Although the method described here is elegant and totally
model-independent, it may fail if  the key asymmetries turn out to be too
small to measure precisely. In this section, we present an alternative
method which depends only on measuring $A_N$ and $A_N'$, asymmetries which
will almost certainly be large enough to measure accurately. The price to
pay will be that some model dependence will enter, but this dependence can
probably be controlled and corrected for \cite{BZK1,bs2, KT}. 

To a first approximation, the amplitude for flipping the spin of ${}^3$He is
the same as that for flipping the neutron spin in $pn$ elastic scattering.
So we begin with the simple parametrization of the pn amplitudes that
corresponds to that used previously for the $pp$ amplitudes \cite{bigpaper}.
\bea
\phi_+(s,t) &=& {s\over 8 \pi}\, \sigma_{\rm tot}^{pn} \, (i +\rho^{pn})
e^{B^{pn}\,t/2}, \nonumber \gp
\phi_5(s,t) &=& \frac{\tau_p \sqrt{-t}}{2 m}\, \phi_+(s,t), \nonumber \gp
\phi_6(s,t) &=& -\frac{\tau_n \sqrt{-t}}{2m}\, \phi_+(s,t).
\eea
 We will neglect $\phi_-$, $\phi_2$ and
$\phi_4Ê$. See
\cite{bigpaper} for a discussion of this approximation. In general, $\tau_p$
and
$\tau_n$ are complex, but Regge arguments favor nearly real
values for high energy. (There is a factor of 2 different in the convention
for $\tau$ from \cite{bigpaper}. This is chosen to parallel more closely
the anomalous magnetic moment in the electromagnteic amplitudes.)

We will now make several approximations in order to keep the discussion
simple. They can be corrected for if a more precise calculation is needed
to interpret experiment. First, we assume that we are at sufficiently high
energy that the scattering is given by pure $I=0$ exchange. This is almost
certainly true at RHIC but corrections can be made when using the approach at
lower energy \cite{berger}. Thereby we will have $\tau_p = \tau_nÊ\equiv \tau$ and
the other parameters $\sigma_{\rm tot}^{pn}, \rho^{pn}, B^{pn}$ will have the
values of the corresponding $pp$ parameters. Next, we will assume
independent particle harmonic oscillator wave functions for the nucleons in
${}^3 {\rm He}$ with the two protons coupled to spin singlet so all the spin is
carried by the neutron. Finally, we will use the impulse approximation for
calculating the $p\, {}^3 {\rm He}$ scattering in terms of $pn$ and $pp$ scattering.
These aspects can easily be improved upon by using better wave functions
and multiple scattering theory \cite{KT}. The results here should not be
misleading at the very small $t$-values of the CNI peak but the corrections should
be calculated if data at larger values of $|t| \geq 0.1 \, GeV^2$ are used in
fitting data to the calculated curve.

Thus we obtain for the hadronic part of the $p\, {}^3 {\rm He}$ amplitudes

\bea
\phi_+^{\,p \,{}^3 {\rm He}}(3s,t) &=& {9 s \over 8 \pi}\, \sigma_{\rm tot}(s) \, (i
+\rho) F_H(t), \nonumber \gp
\phi_5^{\,p\, {}^3  {\rm He}}(3s,t) &=& \frac{\tau \sqrt{-t}}{2m}\,
\phi_+^{\,p \,{}^3 {\rm He}}(3s,t),
\nonumber
\gp
\phi_6^{\,p\, {}^3 {\rm He}}(3s,t) &=& -\frac{1}{3}\frac{\tau \sqrt{-t}}{2m}\,
\phi_+^{\,p \,{}^3 {\rm He}}(3s,t).
\eeaÊ

Note here that $s$ denotes the $s$-value for the $pn$ or $pp$ subsystem and 
$\sigma_{\rm tot}$ denotes the $pp$ total cross-section, so $\sigma_{\rm
tot}^{p\, {}^3 {\rm He}}(3 s) = (8 \pi/3 s) \im{\phi_+}(3s,0)$ . The
factor 3 in the first equation (and the corresponding 1/3 in the last)
results from there being three nucleons in
${}^3 {\rm He}$ (but only one neutron).
$F_H(t) =
\exp\{t (B/2 + a^2/4)\}$ where $a^2$ is the harmonic
oscillator parameter, approximately equal to 57.4 $GeV^{-2}$.

In order to calculate the CNI effect, we need also the electromagnetic
amplitudes for $p \,{}^3 {\rm He}$ scattering. These are (we suppress an index
referring the $p\, {}^3 {\rm He}$ for typographical clarity)

\bea
\phi_+^{em}(3s,t) &=&  \frac{6 s\, \alpha}{t}\, F_{em}(t), \nonumber \gp
\phi_5^{em}(3s,t) &=& -\frac{6 s\, \alpha}{2 m \sqrt{-t}}\, \kappa_p F_{em}(t)
,
\nonumber
\gp
\phi_6^{em}(3s,t) &=& \frac{3 s\,\alpha}{2 m \sqrt{-t}}\, \kappa_n
F_{em}(t) ,
\eea
where $F_{em}(t)=\exp{(a^2\,t/4)}$.Ê
Note that $m$ in all these formulas denotes the proton mass. In the second of
these equations one might want to use the magnetic moment of ${}^3 {\rm He}$ rather
than
$\kappa_n$; this is about $10\%$ larger in magnitude. We leave the
expressions this way for consistency with our simple approach.

Using these relations one can calculate the quantities $a_N, a_{N}\,', b_N,
b_{N}\,'$ as well as $a_0$ and $b_0$. Normalizing to the total cross section
$ A \,\sigma_{\rm tot}$ where $A$ is the nuclear number we get
\bea
a_0 & = & Z\rho, \gp
b_0 & = & \half A \,\lp 1 +\rho^2 \rp ,  \gp
a_N & = & \half P\,Z\,\lb \kappa_p - {\rm Re}(\tau) - \rho \,{\rm Im}(\tau)
\rb
,
\label{eq:a_N}
\gp b_N & = & \half P\, A \, {\rm Im}(\tau) \,\lp 1 +\rho^2 \rp, \gp
a_{N}\,' & = & \half P' \,\lp \kappa_n - {Z \over A} \,\lb{\rm Re}(\tau)
+\rho
\,{\rm Im}(\tau)\rb \rp, \label{eq:a_{N'}} \gp 
b_{N}\,' & = & \half P' \, {\rm Im}(\tau) \,\lp 1 + \rho^2 \rp.
\eea
Here, of course, $A=3$ and $Z=2$; we write the expressions in this way so
that the source of the factors which differ from the $pp$ case is
transparent.

The important result that $b_N$ and $b_{N'}$ are simply related to each
other results from the assumption of $I=0$ dominance of both the flip and
non-flip amplitudes. If one knew the proton polarization $P$ from an
independent polarimeter, then this simple relation immediately gives $P'$.
It is more in the spirit ot the present work to use this one experiment to
determine everything.  Thus
\be
\frac{b_N}{b_{N}\,'} = A\, {P \over P'} \equiv {\mathcal M}_{2}.
\ee
This is to be used then with 
\be
{a_N \over a_{N}\,'} = \frac{Z\,\lb \kappa_p - {\rm Re}(\tau) - \rho \, {\rm
Im}(\tau)\rb}{\kappa_n - {Z \over A} \,\lb{\rm Re}(\tau) +\rho \,{\rm
Im}(\tau)\rb} \, {P \over P'} \equiv {\mathcal M}_{1}
\ee
to solve for ${\rm Re} (\tau [1 - i \rho])$ in terms of these two measurable
ratios ${\mathcal M}_{1}$ and ${\mathcal M}_{2}$:
\be
{Z \over A} {\rm Re} (\tau \lb 1 - i \rho\rb) = \frac{\kappa_n {\mathcal
M}_{1} - {Z \over A}\, \kappa_p {\mathcal M}_{2}}{ {\mathcal M}_{1} -
{\mathcal M}_{2}}.
\ee
With this result in hand, one can use $a_N$ to obtain $P$ and $a_{N}\,'$ to
obtain $P'$. The only fly in the ointment---and it is a pretty serious
fly---is that if ${\rm Im}(\tau)$ should vanish the method fails at step
one because both $b_N$ and $b_{N}\,'$ then vanish. Regge theory suggests that
this is likely to happen at sufficiently high energy, but at RHIC there may remain a
sufficient phase difference between the flip and non-flip amplitudes that both $b_N$
and $b_{N}\,'$ are measurably finite. If ${\rm Im}(\tau)$ does vanish, then this
method can be used as a polarimeter for
${}^3 {\rm He}$ in conjunction with an independent measurement of
the proton beam polarization $P$: for then one can use
Eq.~(\ref{eq:a_N}) to obtain $\tau$ and use it in Eq.\ (\ref{eq:a_{N'}}) to
obtain $P'$.

\section {Conclusions}
This paper presents a model independent method for determining the polarization of
colliding spin
$1/2$ particles, like or unlike, based on measuring the various single- and
double-spin asymmetries using both longitudinal and transversely polarized beams.
No independent information regarding the polarization is needed. The method is
dependent on precise measurements of several probably very small quantities, and
this may limit its applicability, but the existence of the method is interesting
in itself. Finally, an alternative method is given, which depends on some
nuclear physics and some high energy approximations, but which should almost
certainly be practical for unlike particles.

 Acknowledgements: The authors thank J. Millener, B. Kopeliovich and J. Soffer for
valuable discussions. N.H.B. is grateful to Enterprise Ireland for partial support
       provided under
    International Collaboration Programmes IC/1999/075 and IC/2000/027. E.L. is
grateful to the Leverhulme Trust for an Emeritus Fellowship. This
manuscript has been authored under contract number DE-AC02-98CH10886 with
the U.S. Department of Energy. Accordingly, the U.S.  Government retains a
non-exclusive, royalty-free license to publish or reproduce the published
form of this contribution, or allow others to do so, for U.S. Government
purposes.

\end{document}